\documentclass[a4paper,11pt]{article}
\usepackage{pos}

\title{P371 Experiment at CERN - quest for polarized antiprotons}
\author*[a]{M.~Zieliński}
\author[b,c]{D.~Grzonka}
\author[f]{G.~Khatri}
\author[e]{P.~Kulessa}
\author[b,d,c]{J.~Ritman}
\author[b]{T.~Sefzick}
\author[a]{J.~Smyrski}
\author[d,b]{V.~Verhoeven}
\author[b]{H.~Xu}
\onbehalf{\\the P371 Collaboration}

\affiliation[a]{M. Smoluchowski Institute of Physics, Jagiellonian Univeristy, 30-348 Kraków, Poland}
\affiliation[b]{GSI Helmholtzzentrum f\"{u}r Schwerionenforschung GmbH, 64291 Darmstadt, Germany} 
\affiliation[c]{Institute of Nuclear Physics, Forschungszentrum J\"{u}lich, 52428 Jülich, Germany} 
\affiliation[d]{Ruhr-Universit\"{a}t Bochum, Institut f\"{u}r Experimentalphysik I, 44801 Bochum, Germany} 
\affiliation[e]{H.~Niewodniczański Institute of Physics, Polish Academy of Science, 31-342 Kraków, Poland}
\affiliation[f]{CERN, 1211 Geneva 23, Switzerland}

\emailAdd{marcin.zielinski@uj.edu.pl}

\abstract{Polarization effects in the production of antiprotons at the CERN PS beam line T11 at 3.5 GeV/c have been investigated within the P371 experiment. These effects, if found to be significant could provide a simple method to generate polarized antiproton beams with existing facilities. First precursor measurements were carried out by the P349 collaboration, though the available statistics were insufficient for a quantitative conclusion. With an upgraded detector setup and extended beam time, the experiment aims at determining whether a measurable degree of antiproton polarization exists.}

\FullConference{The 21st International Conference on Hadron Spectroscopy and Structure (HADRON2025)\\
 27 - 31 March, 2025\\
Osaka University, Japan\\}

\begin{document}
\maketitle

\section{Introduction}
Spin degrees of freedom play a central role in hadronic physics. Observables involving polarization are particularly sensitive to the underlying dynamics and often provide information that cannot be accessed through unpolarized measurements. Ideally, a polarized beam in combination with a polarized target allows one to prepare well defined quantum states of the interacting system. In the case of $\bar{p}p$ reactions, a configuration with parallel spins ($\bar{p}^{\uparrow}p^{\uparrow}$) corresponds to a pure triplet state, while in the antiparallel case ($\bar{p}^{\uparrow}p^{\downarrow}$) the singlet component dominates. Controlling such spin configurations is therefore of great importance in a wide range of studies, from precision tests at low energies to investigations of fundamental QCD processes~\cite{KLEMPT2002119}. 

In contrast to protons, for which polarized beams and targets are routinely available, the case of antiprotons still remains unsettled. Various concepts for generating polarized antiproton beams have been explored since cooled antiproton facilities have been routinely used, e.g. polarization transfer from hyperon decay, spin filtering and spin–flip processes, synchrotron–radiation effects, interactions with polarized photons, the Stern–Gerlach effect~\cite{10.1063/1.35683,2008AIPC.1008..124M,2009AIPC.1149...80S}. In many of the suggested approaches, the expected number of polarized antiprotons or the achievable polarization is far too small, and in several cases, reliable estimates cannot even be made, which has so far discouraged detailed feasibility studies. The question of how to obtain a polarized antiproton beam therefore remains open and continues to motivate experimental studies of P371 Collaboration. 

A conceptually simple path towards a polarized antiproton beam is the production process itself. The idea that antiprotons might already acquire a polarization at the moment of their production motivated the proposal of the P349 experiment~\cite{Kilian:2011zz,Grzonka:1697756,Grzonka:2015bep}. First  measurements were carried out at CERN in 2014, 2015, and 2018 but due to limited statistics, no conclusions could be drawn. Based on this precursor study, in 2025 an upgraded detector setup together with extended beam time has been prepared  in order to reach the accuracy required to determine whether a measurable degree of polarization is indeed present in the production mechanism. In these proceeding we  describe short physics motivation and general detector setup, followed by the conclusions and outlook. 

\section{Polarization in antiprotons production}
Antiprotons are produced in proton–nucleon interactions when a high energy proton beam collides with a fixed target. At the Proton Synchrotron (PS) at CERN, protons with a momentum of 24~GeV/$c$ are used, yielding approximately one antiproton per $10^{6}$ incident protons. The dominant production channel is a quasi-free interaction $pp\rightarrow pp\bar{p}p$, and the resulting antiproton momentum spectrum reaches a maximum near 3.5~GeV/$c$. If this production process would lead to transverse polarization of antiprotons, a polarized beam could in principle be obtained by selecting particles from one side of the angular distribution, while excluding the forward region dominated by the $S$-wave ($\theta < 50$~mrad). Such a selection could be realized with comparatively simple modifications of an antiproton production and cooling line by introducing appropriate absorbers. 

In order to determine the polarization of the produced antiprotons, it is necessary to employ a secondary scattering process with well known and sufficiently large analyzing power $A_{y}$. This is valid for protons in elastic $pp$ scattering within the Coulomb–nuclear interference (CNI) region. At high energies, the non zero analyzing power in this kinematic region arises from the interference between the non spin flip nuclear amplitude and the electromagnetic spin flip amplitude~\cite{FNAL-E581704:1989hno,E581704:1993mgc,OKADA2006450}. A maximum analyzing power of about 4.5\% is reached for a four-momentum transfer at the peak $t_p$ given by~\cite{E581704:1993mgc}:
\[t_p = -\frac{8\pi\sqrt{3}\,\alpha}{\sigma_{\mathrm{tot}}},\] where $\alpha$ denotes the fine-structure constant and $\sigma_{\mathrm{tot}}$ the total cross section, when  assuming $\sigma_{\mathrm{tot}} = 40$~mb. Experimentally, a maximum $A_y$ of about 4.5\% was observed at $t = 0.0037~(\mathrm{GeV}/c)^{2}$~\cite{OKADA2006450}. For antiprotons, the analyzing power in the CNI region is expected to be of the same magnitude as for protons, since the hadronic contribution reduces to the non spin flip amplitude while the Coulomb spin flip term only changes sign. This behavior was confirmed experimentally at Fermilab with 185~GeV/$c$ polarized antiprotons, where an analyzing power of about $(-4.6 \pm 1.9)$\% was observed~\cite{E581704:1989vcm}. At 3.5~GeV/$c$, and a laboratory scattering angle of roughly 20~mrad, a similar analyzing power of about 4.5\% is anticipated.

\section{Experiment and detector design}
The study of antiproton polarization will be done at the T11 beam line of the CERN PS, which provides favorable conditions for this investigation. Secondary particles are selected at an angle of about 150~mrad, and the beam line can be tuned to momenta around 3.5~GeV/$c$ for both positively and negatively charged particles. For negatively charged particles at 3.5~GeV/$c$, the momentum resolution is about 5\%, giving roughly $10^{6}$ particles per spill of 400~ms length. The yield of antiprotons at T11 is expected to be in the order of $8 \times 10^{3}$ $\bar{p}$ per spill. 

\begin{figure}[t]
\centerline{\includegraphics[width=13.25cm]{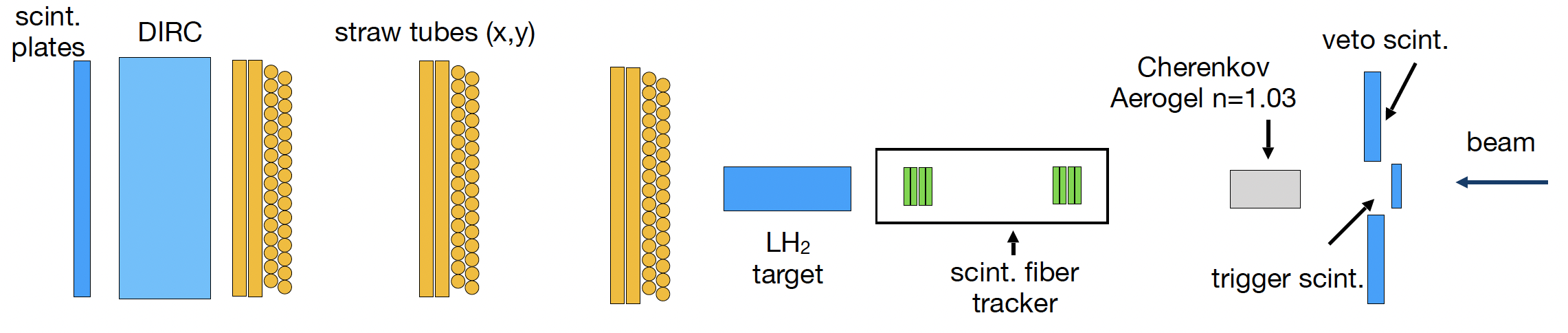}}
\caption{Schematic view of the detector arrangement in the horizontal plane. 
The beam enters from the right and passes sequentially through the start scintillator, Cherenkov detector, scintillating fiber hodoscope, scattering target, straw tubes, DIRC, and scintillator plates.}
\label{Fig:f1}
\end{figure}

The experimental setup is designed to provide efficient triggering, tracking, and particle identification, while minimizing multiple scattering effects. It consists of scintillators for trigger generation, an aerogel Cherenkov detector for online pion suppression, scintillating fibers for tracking of produced antiprotons, a liquid hydrogen analyzer target (LH2 target), straw tubes to reconstruct the trajectories of primary and scattered antiprotons, and a DIRC detector for offline particle identification. For the data acquisition, a DOGMA system will be used~\cite{dogma}. The detection system is operated in air, and the overall contribution from material effects, including straggling, remains below the intrinsic angular resolution of the tracking detectors (about 1~mrad for the fiber system and below 0.5~mrad for the straw tubes). A schematic view of the detector layout is presented in Fig.~\ref{Fig:f1}.

Based on the detector configuration and beam conditions to evaluate the feasibility of the polarization measurement, the expected yield of elastic $\bar{p}p$ scattering  events in the relevant kinematic range has been estimated. The integral of the $\bar{p}p$ cross section in the range $t=-0.002$ to $-0.007$~(GeV/$c$)$^{2}$ is about 1.35~mb~\cite{E760:1996mur}, which corresponds to the region of sufficiently large $A_y$. With an estimated yield of $8\times 10^{3}$ antiprotons per spill and a 12~cm thick liquid hydrogen target, about seven useful scattering events per spill are expected. With an expected mean of 4000 spills per day, approximately $1.6\times 10^{6}$ events are collected over 8 weeks of beam time. Simulation studies indicate that such statistics are 
sufficient to measure a statistically non-zero asymmetry for polarization values of a few percent.

\section{Summary}
The investigation of polarization effects in antiproton production offers a promising possibility for the generation of polarized antiproton beams with existing accelerator facilities. For measuring the antiproton polarization, elastic $\bar{p}p$ scattering in the CNI region is employed as a well-understood analyzing reaction. 
A dedicated setup combining tracking and timing detectors, an analyzer target, and particle identification systems has been prepared and will be operated at the CERN T11 beam line. 
With the expected beam conditions, a data sample of the order of $10^{6}$ $\bar{p}p$ scattering events can be collected in a few weeks of operation, sufficient to extract an asymmetry with sensitivity to polarization values at the few-percent level. The forthcoming measurements will thus allow, for the first time, a direct test of whether the antiproton production process itself generates polarization.

%%%%%%%
\section*{Acknowledgments}
This work was supported partially by the "Research Support Module" as part of the "Excellence Initiative - Research University" program at the Jagiellonian University in Kraków. 

%%%%%%%
\bibliographystyle{unsrt}
\bibliography{bib2}

\end{document}